# Response of a Commercial 0.25 μm Thin-Film Silicon-on-Sapphire CMOS Technology to Total Ionizing Dose

**Michael P. King**[a]*****, **Datao Gong**[b], **Chonghan Liu**[b], **Tiankuan Liu**[b], **Annie C. Xiang**[b], **Jinbo Ye**[b], **Ronald D. Schrimpf**[a], **Robert A. Reed**[a], **Michael L. Alles**[a], **and Daniel M. Fleetwood**[a]

[a] *Vanderbilt University,*
  *Nashville, TN 37235, USA*
  *E-mail:* michael.p.king@vanderbilt.edu

[b] *Southern Methodist University,*
  *Dalls, TX 75205, USA*

ABSTRACT: The radiation response of a 0.25 μm silicon-on-sapphire CMOS technology is characterized at the transistor and circuit levels utilizing both standard and enclosed layout devices. Device-level characterization showed $\Delta V_T$ of less than 170 mV and $\Delta I_{LEAKAGE}$ of less than 1 nA for individual nMOSFET and pMOSFET devices at a total dose of 100 krad(SiO$_2$). The increase in power supply current at the circuit level was less than 5%, consistent with the small change in off-state transistor leakage current. The technology exhibits good characteristics for use in the electronics of the ATLAS experiment at the Large Hadron Collider.

---

***** Corresponding author.

# Contents

## 1. Introduction

Silicon-on-sapphire (SoS) complementary metal-oxide-semiconductor (CMOS) technology has been used in radiation-tolerant applications since the 1970s. SoS technologies exhibit several characteristics that make them attractive for use in radiation environments, including an insulating sapphire layer below the active silicon that eliminates the parasitic inter-device bipolar structure associated with latchup in bulk devices. SoS technologies also have been reported to have smaller single-event upset cross-sections than equivalent bulk processes [1]. However, radiation-induced leakage currents along the edges of the device and the back channel, where the active silicon meets the sapphire substrate, are important issues in these technologies [2]. The ATLAS experiment at the Large Hadron Collider is one example of an application for which SoS technology is very promising; the radiation environment is quite large compared to typical space and defense applications [3].

In this work, the radiation response of a 0.25 μm SoS CMOS technology is characterized at the transistor and circuit levels. Devices are evaluated in both standard and enclosed layout geometries. Device-level characterization shows radiation-induced charge trapping in the gate oxide results in threshold voltage shifts less than 170 mV to a total dose of 100 krad($SiO_2$). Additionally, increases in radiation-induced leakage current are less than 1 nA for both nMOSFET and pMOSFET devices. Circuit-level evaluation of these structures is consistent with these results.

## 2. Test Structures and Experimental Conditions

The Peregrine 0.25 μm SoS process is a thin-film technology with both partially depleted and fully depleted devices. The epitaxial silicon layer is 80 nm thick with a 200 μm sapphire insulating substrate. The gate oxide thickness is 6 nm, and the process uses LOCOS for device isolation. Individual transistors and a set of shift registers were fabricated on a test chip.

Structures were fabricated in either standard or enclosed layout geometries. Three different transistor types were used with high, regular, or intrinsic threshold voltages. Devices with regular and high threshold voltages are partially depleted, while the intrinsic devices are fully depleted. Devices were packaged and subsequently baked for twelve hours at 150°C in preparation for irradiation. nMOSFET and pMOSFET irradiation bias conditions were

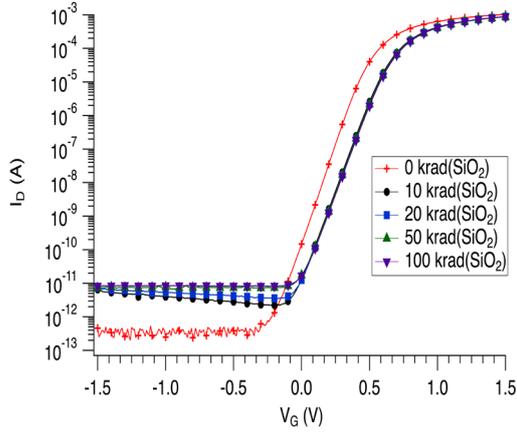 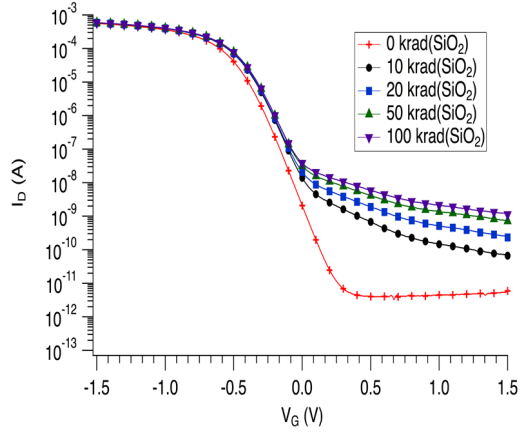

Fig 1a $I_D$-$V_G$ characteristics of an enclosed layout, regular threshold voltage nMOS SOS transistor.

Fig. 1b $I_D$-$V_G$ characteristics of an enclosed layout, regular threshold voltage pMOS SOS transistor.

$V_D$ = 2.5 V and -2.5 V, respectively; all other terminals were grounded. This off-state condition is the worst case for inverters exposed to total ionizing dose [4]. The transistors were irradiated with 10 keV x-rays at a dose rate of 31.5 krad(SiO$_2$) per minute using an ARACOR Model 4100 irradiator. $I_D$-$V_G$ sweeps were performed to characterize the leakage current and threshold voltage of irradiated devices. Device characterization was performed with an HP 4156A parameter analyzer with an applied drain bias of ±0.1 V for nMOS and pMOS transistors respectively. Gate voltages were swept between -1.5 V to 1.5 V. The source and sapphire substrate were grounded during device characterization and irradiation.

Two types of shift registers, consisting of 32 D-flip-flop stages, one using standard layout transistors, the other enclosed layout transistors, were fabricated for circuit-level evaluation of radiation-induced leakage current. Power supply to the shift registers was 2.5 V with the substrate grounded. The shift registers were irradiated with 198 MeV protons to a total fluence of $1.27 \times 10^{13}$ cm$^{-2}$. The operating frequency during irradiation was 40 MHz; the power supply current for each shift register was monitored with a Keithley 2700 multi-channel digital multimeter.

## 3. Experimental Results

Typical pre-rad and post-rad $I_D$-$V_G$ characteristics are shown in Figs. 1a and 1b for nMOS and pMOS transistors. These devices are enclosed layout, regular threshold voltage devices, corresponding to $V_T$ = 0.55 V and -0.35 V, for nMOS and pMOS transistors, respectively.

An initial positive shift in threshold voltage, with a maximum value of 170 mV, was observed at less than 1 krad(SiO$_2$) in both nMOSFET and pMOSFET devices, as seen in Fig. 2. This initial shift is due to radiation-induced electron trapping in the sapphire substrate [5]. These trapped electrons accumulate the back channel of nMOSFETs and the front to back coupling results in a shift in threshold voltage. In pMOSFETs the trapped electrons couple the front and back gate similarly, however, they deplete the n-type body, leading to additional leakage current. No additional threshold-voltage shifts were observed following the initial exposure of 10 krad(SiO$_2$), up to the largest dose considered here (100 krad(SiO$_2$)).

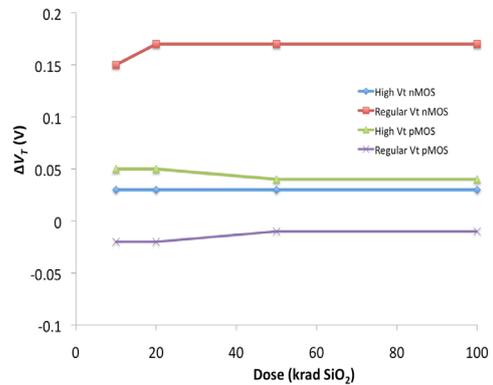

Fig. 2 $\Delta V_T$ as a function of dose for partially depleted nMOS and pMOS devices.

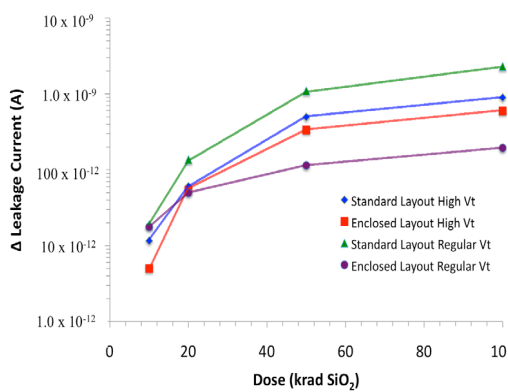 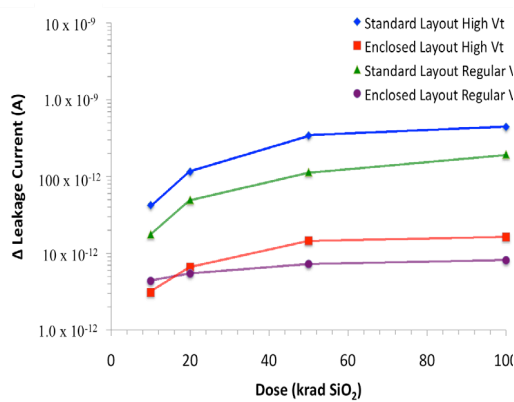

Fig. 3a Change in leakage current with dose for high and regular threshold pMOSFETs in standard and enclosed layout.

Fig. 3b Change in leakage current with dose for high and regular threshold nMOSFETs in standard and enclosed layout.

The radiation-induced leakage current for standard and enclosed layout devices is shown in Figs. 3a and 3b for pMOSFETs and nMOSFETs, respectively. nMOS transistors in a standard layout configuration exhibit a parasitic conductive path along the edge of the device [6]. The enclosed layout transistors eliminate the edge leakage paths present in the standard layout nMOS devices. The edge leakage is associated with hole trapping in the isolation oxide, which affects nMOS transistors. Conversely, leakage paths exist for pMOS transistors primarily along the back channel of the device due to electron trapping, which impacted the threshold voltage of both nMOS and pMOS transistors. This back-channel leakage path exists and impacts both standard and enclosed layout device geometries.

The post-irradiation increase in power supply current at the circuit level is less than 5%, which is consistent with the relatively small radiation-induced change in off-state transistor leakage (less than 1 nA at 100krad(SiO$_2$) for both nMOSFET and pMOSFET devices). Leakage current was consistently higher for standard layout devices than for enclosed layout devices following irradiation because of the elimination of the parasitic edge leakage path in the enclosed layout devices. The increase in circuit-level leakage current with increasing dose is caused by the formation of a back channel in the pMOS devices along the sapphire-silicon interface.

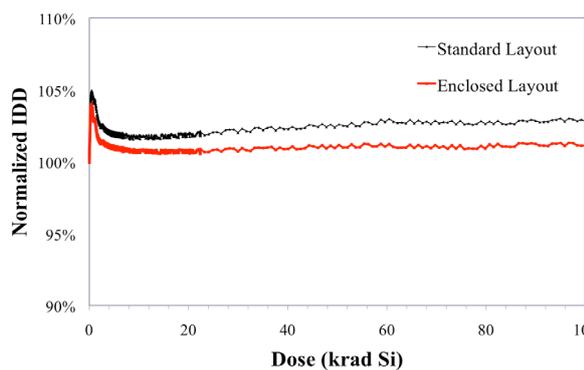

Fig. 5 Normalized power supply current of standard and enclosed layout shift registers irradiated with 198 MeV protons.

## 4. Conclusions

Single transistors and shift registers fabricated in a 0.25 μm SoS CMOS technology were irradiated with 10 keV x-rays and 198 MeV protons, respectively. Radiation-induced electron trapping at the silicon-sapphire interface results in a shift in threshold voltage for both nMOS and pMOS transistors. The overall magnitude of the shifted threshold characteristics was less than 170 mV, and saturated within 1 krad($SiO_2$). At the transistor level, pMOSFETs in a standard layout exhibited the largest increases in leakage current due to the of the back-channel. Radiation-induced leakage current was observed to be 1 nA or less. This result is consistent with circuit-level results during proton irradiation of shift registers, and had little impact on circuit operation. These results indicate 0.25 μm SoS exhibits stable operating characteristics in a total dose environment of 100 krad($SiO_2$) or less, and appears to be very well suited for operating in the ATLAS TID environment.

## Acknowledgments

The authors would like to thank Peregrine Semiconductor. This work is supported by US Department of Energy Grant DE-FG02-04ER41299.